\documentclass[conference]{IEEEtran}
\usepackage{hyperref}
\usepackage[noadjust]{cite}
\usepackage[cmex10]{amsmath}
\usepackage[caption=false,font=footnotesize]{subfig}
\usepackage{url}
\usepackage{graphicx}
\usepackage{amssymb}
\usepackage{centernot}
\everymath{\displaystyle}
\usepackage{comment}

\usepackage[shortlabels]{enumitem}
\usepackage{nameref}
\usepackage{slashbox}
\usepackage{textcomp}
\usepackage{gensymb}
\usepackage[all]{hypcap}
\usepackage{tikz}
\usetikzlibrary{patterns}

\DeclareMathOperator{\round}{round}
\DeclareMathOperator{\functioncheck}{check}
\DeclareMathOperator{\functionangle}{angle}

\mathchardef\UrlBreakPenalty=10000
\mathchardef\UrlBigBreakPenalty=10000

\begin{document}
\title
{
  Searching for a Compressed Polyline\\
  with a Minimum Number of Vertices
}
\author
{
\IEEEauthorblockN{Alexander Gribov}
\IEEEauthorblockA
{
  Environmental Systems Research Institute\\
  380 New York Street\\
  Redlands, CA 92373\\
  E-mail: agribov@esri.com}
}

\maketitle

\begin{abstract}
\boldmath
  There are many practical applications that require simplification of polylines. Some of the goals are to reduce the amount of information necessary to store, improve processing time, or simplify editing. The simplification is usually done by removing some of the vertices, making the resultant polyline go through a subset of the source polyline vertices. However, such approaches do not necessarily produce a new polyline with the minimum number of vertices. The approximate solution to find a polyline, within a specified tolerance, with the minimum number of vertices is described in this paper.
\end{abstract}


\section{Introduction}

The task is to find a polyline, within a specified tolerance of the source polyline, with the minimum number of vertices. That polyline is called optimal. Usually, a subset of vertices of the source polyline is used to construct an optimal polyline~\cite{CompressionAlgorithm, CompressionReview}. However, an optimal polyline does not necessarily have vertices coincident with the source polyline vertices. One approach, to allow the resultant polyline to have flexibility in the locations of vertices, is to find the intersection between adjacent straight lines~\cite{PolylineGeneralizationCombinatorical} or geometrical primitives~\cite{PolylineGeneralization}. However, there are situations when such an approach does not work well, for example, when adjacent straight lines are almost parallel to each other or a circular arc is close to being tangent to a straight segment. The approach described in this paper evaluates a set of vertex locations (considered locations) while searching for a polyline with the minimum number of vertices.

\section{Algorithm}

\subsection
{
  Discretization of the Solution
  \label{sec:Descretization}
}

\begin{figure} [b]
  \centering
  \includegraphics[width = 6 cm, keepaspectratio]{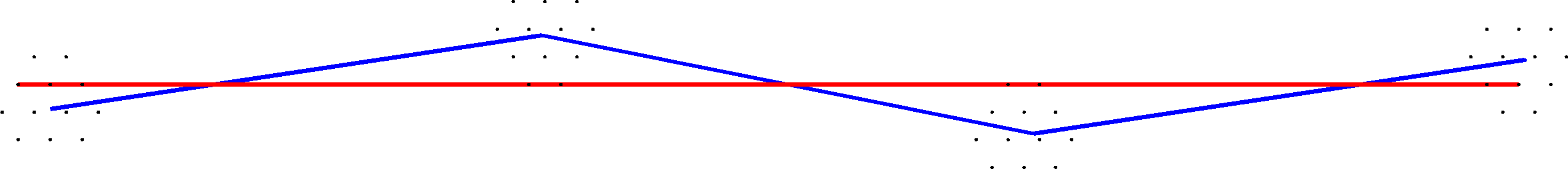}
  \caption
  {
    Example of one segment (red segment) between considered locations (black dots) within tolerance of the source polyline (blue polyline).
  }
  \label{fig:ExampleSegment}
\end{figure}

Any compressed polyline must be within tolerance of the source polyline; therefore, the compressed polyline must have vertices within tolerance of the source polyline. It would be very difficult to consider all possible polylines and find one with the minimum number of vertices; therefore, as an approximation, only some locations around vertices of the source polyline are considered (see the black points around the vertices of the source polyline in Fig.~\ref{fig:ExampleSegment}).

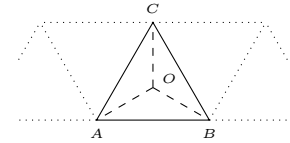
\begin{figure} [htb]
  \centering
  \begin{tikzpicture}[scale = 1.5]
    \tikzstyle{every node} = [font = \tiny]

    \begin{scope}
      \clip (-0.7, -0.2) -- (1.7, - 0.2) -- (1.7, 0.86602540378443864676372317075294 + 0.2) -- (-0.7, 0.86602540378443864676372317075294 + 0.2) -- cycle;
      \draw [dotted] (0, 0) -- (-1, 0);
      \draw [dotted] (1, 0) -- (2, 0);
      \draw [dotted] (0.5, 0.86602540378443864676372317075294) -- (-0.5, 0.86602540378443864676372317075294);
      \draw [dotted] (0.5, 0.86602540378443864676372317075294) -- (1.5, 0.86602540378443864676372317075294);
      \draw [dotted] (0, 0) -- (-0.5, 0.86602540378443864676372317075294);
      \draw [dotted] (1, 0) -- (1.5, 0.86602540378443864676372317075294);
      \draw [dotted] (-1, 0) -- (-0.5, 0.86602540378443864676372317075294);
      \draw [dotted] (2, 0) -- (1.5, 0.86602540378443864676372317075294);
    \end{scope}
    
    \draw (0, 0) -- (1, 0) -- (0.5, 0.86602540378443864676372317075294) -- cycle;
    \draw [dashed] (0.5, 0.28867513459481288225457439025098) -- (0, 0);
    \draw [dashed] (0.5, 0.28867513459481288225457439025098) -- (1, 0);
    \draw [dashed] (0.5, 0.28867513459481288225457439025098) -- (0.5, 0.86602540378443864676372317075294);
    \draw (0.5, 0.28867513459481288225457439025098) node [anchor = 210] {$O$};
    \draw (0, 0) node [anchor = 90] {$A$};
    \draw (1, 0) node [anchor = 90] {$B$};
    \draw (0.5, 0.86602540378443864676372317075294) node [anchor = 270] {$C$};
  
  \end{tikzpicture}
  \caption
  {
    The worst case distance for the equilateral triangular grid is the distance from the center of the triangle $O$ to any vertex of the equilateral triangle. If $OA = OB = OC = 1$, then $AB = BC = CA = \sqrt{3}$.
  }
  \label{fig:TriangularGrid}
\end{figure}

The locations around vertices of the source polyline are chosen to be on an infinite equilateral triangular grid with the distance from vertices of the source polyline less than the specified tolerance. The equilateral triangular grid (see Fig.~\ref{fig:TriangularGrid}) has the lowest number of nodes versus other grids (square, hexagonal, etc.), satisfying that distance from any point to the closest node does not exceed the specified threshold.

The choice for the side of an equilateral triangle in the equilateral triangular grid is calculated from the error it introduces. That error can be expressed as a proportion of the specified tolerance. For example, $q \in \left( 0, 1 \right)$ proportion of the specified tolerance means that the side of the equilateral triangle is equal to $q \sqrt{3}$ times the specified tolerance. This leads to about $\dfrac{2 \pi}{3 \sqrt{3} q^2} \approx \dfrac{1.2}{q^2}$ locations per each vertex. To decrease complexity, some locations might be skipped; if they are considered in neighbor vertices of the source polyline, however, it should be done without breaking the combinatorial algorithm described in section \ref{sec:CombinatorialApproach}. If tolerance is great, it is possible to consider locations around segments of the source polyline. In this paper, to support any tolerance, only locations around vertices of the source polyline are considered. Densification of the source polyline might be necessary to find the polyline with the minimum number of vertices.

\subsection
{
  Testing a Segment to Satisfy Tolerance
  \label{sec:TestingSegmentTolerance}
}

For a compressed polyline to be within tolerance, every segment of the compressed polyline must be within tolerance from the part of the source polyline it describes. To find the compressed polyline with the minimum number of vertices, this test has to be performed many times for all combinations of possible locations of vertices (see Fig.~\ref{fig:ExampleSegment}). \cite{OptimizedCompressionAlgorithm} describes an efficient approach to perform these tests based on the convex hull. If the convex hull is stored as a polygon, the complexity of this task is $O{\left( \log{n} \right)}$, where $n$ is the number of vertices in the convex hull~\cite{OptimizedCompressionAlgorithm}. The expected complexity of the convex hull for the $N$ random points in any rectangle is $O{\left( \log{N} \right)}$, see~\cite{ConvexHullsComplexity}. If the source polyline has parts close to an arc, the size of the convex hull tends to increase. For the worst case, the number of vertices in the convex hull is equal to the number of vertices in the original set.

If there are no lines with thickness of two tolerances covering the convex hull completely, then one segment cannot describe this part of the source polyline. The complexity of this check is $O{\left( n \log{n} \right)}$.

A convex hull for any part of the source polyline is constructed in the same way as in~\cite{OptimizedCompressionAlgorithm}.

\subsection
{
  Testing Segment End Points
  \label{sec:TestingSegmentEndPoints}
}

\begin{figure} [hb]
  \centering
  \begin{tabular}{c c c}
    \includegraphics[height = 1.5 cm, keepaspectratio]{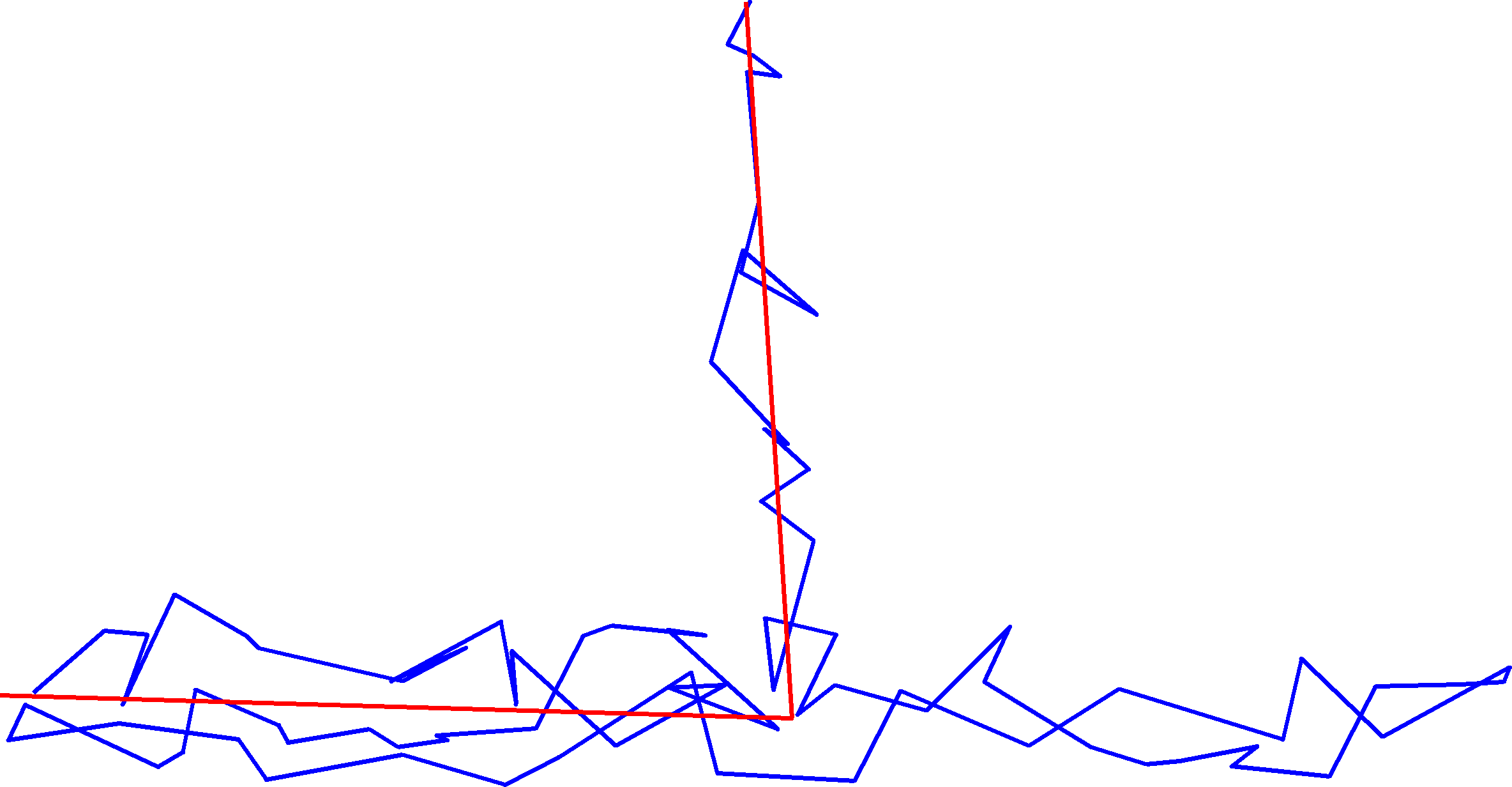} &
    \includegraphics[height = 1.5 cm, keepaspectratio]{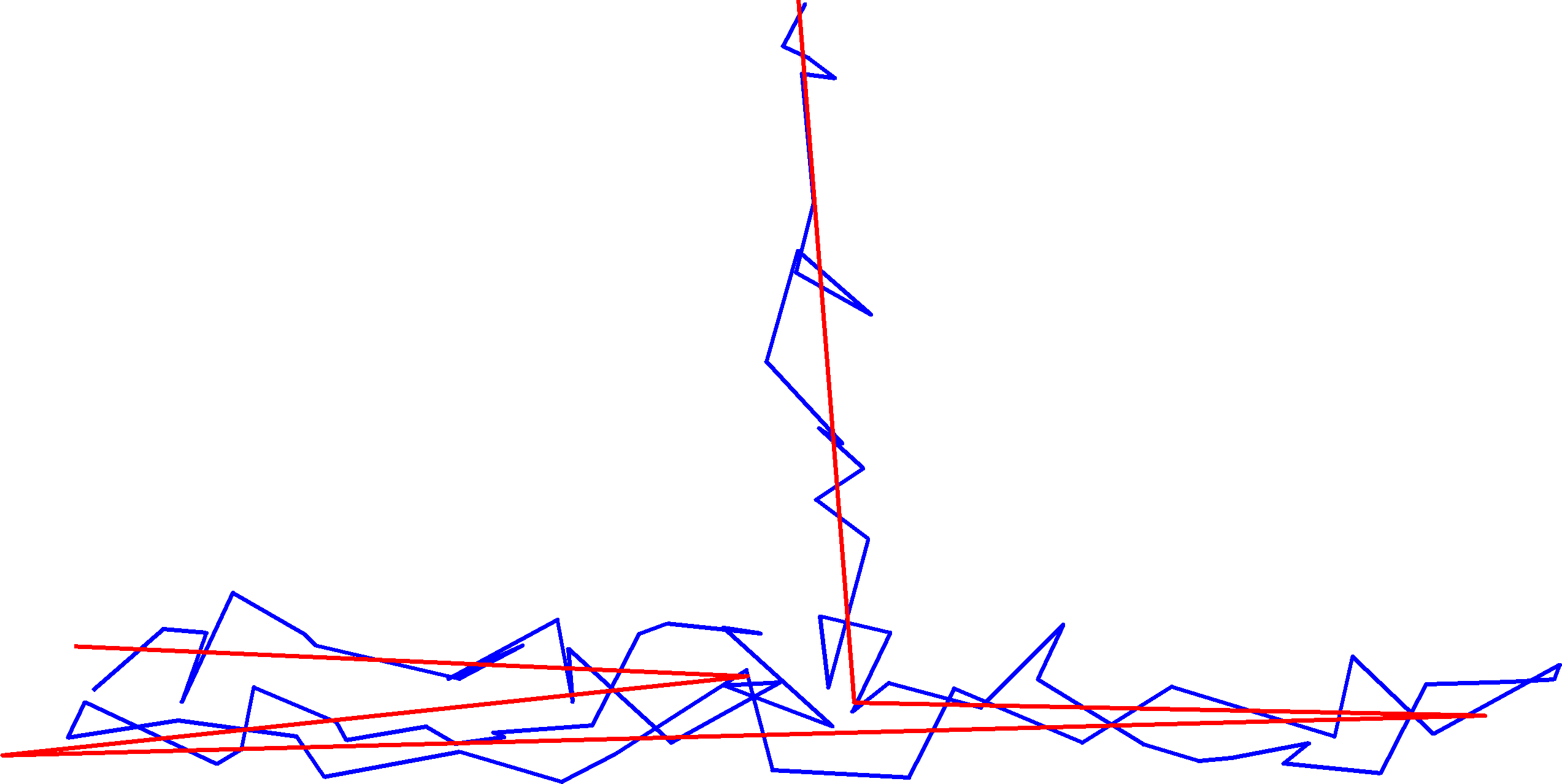} \\
    (a) &
    (b)
  \end{tabular}
  \caption
  {
    The blue polyline is the source polyline. The red polyline is the result of the algorithm without checking for end points and the source polyline direction (a) and with both checks performed (b).
  }
  \label{fig:TopologycalTest}
\end{figure}

The test described in the previous section~\ref{sec:TestingSegmentTolerance} does not check the ends of the segment. The example in Fig.~\ref{fig:TopologycalTest} shows that the source polyline changes direction to the opposite several times (zigzag) before going up. Without checking end points and changes in direction, the compressed polyline might not describe some parts of the source polyline (Fig.~\ref{fig:TopologycalTest}a). Therefore, these tests are necessary to guarantee that the compressed polyline (Fig.~\ref{fig:TopologycalTest}b) describes the source polyline without missing any parts.

The segment end points to be within the tolerance of the part of the source polyline are tested based on the convex hull in the same way as the test for the segment to be within tolerance performed in section~\ref{sec:TestingSegmentTolerance}.

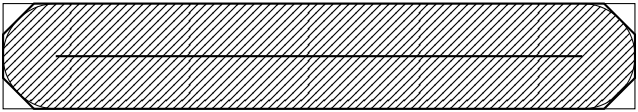
\begin{figure} [htb]
  \centering
  \begin{tikzpicture}[scale = 0.7]
    \draw [thick] (0, 0) -- (10, 0);
    \draw (-1, -1) -- (11, -1) -- (11, 1) -- (-1, 1) -- cycle;
    \draw (0, 1) arc (90:270:1);
    \draw (10, -1) arc (-90:90:1);
    \draw [thick] (-0.4142135623730950488016887242097, 1) -- (-1, 0.4142135623730950488016887242097) -- (-1, -0.4142135623730950488016887242097) -- (-0.4142135623730950488016887242097, -1) -- (10 + 0.4142135623730950488016887242097, -1) -- (11, -0.4142135623730950488016887242097) -- (11, 0.4142135623730950488016887242097) -- (10 + 0.4142135623730950488016887242097, 1) -- cycle;
    \fill [pattern = north east lines] (0, 1) arc (90:270:1) -- (10, -1) arc (-90:90:1) -- cycle;
  \end{tikzpicture}
  \caption
  {
    The diagonal striped area is the tolerance area around the segment. The thin rectangle is the approximation of the area around the segment. A thick polygon would be a better approximation.
  }
  \label{fig:SegmentEndPoints}
\end{figure}

This is equivalent to the test if the segment extended in parallel and perpendicular directions by the tolerance (see Fig.~\ref{fig:SegmentEndPoints}) contains a convex hull of the part of the source polyline it describes. If more directions are used, a better approximation of the curved polygon can be obtained. The complexity of the test is $O{\left( \log{n} \right)}$.

\subsection
{
  Testing Polyline Direction
  \label{sec:TestingZigZag}
}

The test for the source polyline to have a zigzag is performed by checking if the projection to the segment of backward movement exceeds two tolerances ($2 T$, where $T$ is the tolerance). Two tolerances are used because one vertex of the source polyline can shift forward by the tolerance and the vertex after that shift backward by the tolerance. The algorithm is based on analyzing zigzags before the processed point. Let $p_i$ be the vertices of the polyline, $i = \overline{0..N - 1}$, $N$ be the number of vertices in the polyline. The next algorithm constructs a table for efficient testing.
\begin{enumerate}[label={}]
  \item Define a set of directions $\alpha_j = \dfrac{2 \pi}{N_d} j$,\\
  where $j = \overline{0..N_d - 1}$, $N_d$ is the number of directions.
  \item Cycle over each direction $\alpha_j$, $j = \overline{0..N_d - 1}$.
  \begin{enumerate}[label={}]
    \item Define the priority queue with requests containing two numbers. The first number is the real value, and the second number is the index. Priority of the request is equal to the first number.
    \item Set $k = 0$.
    \item Cycle over each point $p_i$ of the source polyline,\\
    $i = \overline{0..N - 1}$.
    \begin{enumerate}[label={}]
      \item Calculate projection of $p_i$ to the direction $\alpha_j$ (scalar product between the point and the direction vector):
      \begin{equation*}
        d = p_i \cdot \left( \cos \left( \alpha_j \right), \sin \left( \alpha_j \right) \right).
      \end{equation*}
      \item Remove all requests from the priority queue with a priority of more than $d + 2 T$. If the largest index from removed requests is larger than $k$, set $k$ equal to that index.
      \item Set $V_{j, i} = k$.
      \item Add request $\left( d, i + 1 \right)$ to the priority queue.
    \end{enumerate}
  \end{enumerate}
\end{enumerate}

To test if the part of the source polyline between vertices $i_s$ and $i_e$ has a zigzag.
\begin{enumerate}[label={}]
  \item First, find the closest direction $\alpha_j$ to the direction of the segment $\alpha_{j^*}$:
  $
    j^* = \round{\left( \dfrac{N_d}{2 \pi} \alpha \right)}
    \!\!\!
    \mod
    N_d
  $,
  where $\alpha$ is the direction of the segment.
  \item Second, if $V_{j^*, i_{e}} \leq i_{s}$, then there are no zigzags for the segment describing the part of the source polyline from vertex $i_{s}$ till $i_{e}$.
\end{enumerate}

Let $W_i = \min_{0 \leq j \wedge j < N_d}{\left( V_{j, i} \right)}$. If $i_{s} < W_{i_{e}}$, then one segment cannot describe the part of the source polyline from vertex $i_{s}$ till $i_{e}$.

This test has some limitations:
\begin{itemize}
  \item The tested direction is approximated by the closest one, making the check approximate.
  \item For some error models, a zigzag might pass the test. For example, if errors are limited by a circle, a zigzag by two tolerances is only possible if it happens directly on the segment.
\end{itemize}

Nevertheless, it is an efficient test to avoid absurd results, like in Fig.~\ref{fig:TopologycalTest}a. The complexity of the algorithm is $O{\left( N_d N \log \left( N \right) \right)}$ and the complexity to test any segment is $O{\left( 1 \right)}$.

\subsection
{
  Combinatorial Approach to Find an Optimal Solution
  \label{sec:CombinatorialApproach}
}

The optimal solution is found by using the algorithm described in~\cite{PolylineGeneralizationCombinatorical}.

Let $p_{i, j}$ be considered locations for vertex $p_i$, where ${i = \overline{0..N - 1}}$, ${j = \overline{0..N_i - 1}}$, $N_i$ is the number of considered locations for the vertex $i$. Let pairs $\left( i_k, j_k \right)$, ${k = \overline{0..m}}$, divide the source polyline into $m$ straight segments
$
  \left( p_{i_k, j_k}, p_{i_{k + 1}, j_{k + 1}} \right)
$
describing the source polyline from vertex $i_k$ till $i_{k + 1}$, $k = \overline{0..m - 1}$. Notice that neighbor segments are already connected in $p_{i_k, j_k}$, $k = \overline{1..m - 1}$, and this solution avoids problems in algorithms~\cite{PolylineGeneralizationCombinatorical, PolylineGeneralization} when the intersection of neighbor segments is far away from the source polyline.

The goal of this algorithm is to find the solution with the minimum number of vertices while satisfying tolerance restriction, and among them with the minimum integral square differences. Therefore, minimization is performed in two parts
$
  \left\{
    \begin{aligned}
      & T^{\#}\\
      & T^{\epsilon}
    \end{aligned}
  \right\}
$,
where the first part $T^{\#}$ is the number of segments, and the second part $T^{\epsilon}$ is the integral of the square deviation between segments and the source polyline. The solutions are compared by the number of segments and, if they have the same number of segments, by square deviation between segments and the source polyline. The solution of this task, when the optimal polyline has vertices coincident with the source polyline, can be found in \cite{CombinatorialMinimumNumberSegments}.

Let $P_k$, $k = \overline{0..N - 1}$ be parts of the source polyline from vertex $0$ to $k$.

The optimal solution is found by induction.
Define the optimal solution for polyline $P_0$ as
$
  \left\{
    \begin{aligned}
      & T_{0, j}^{\#}\\
      & T_{0, j}^{\epsilon}
    \end{aligned}
  \right\}
  =
  \left\{
    \begin{aligned}
      0\\
      0
    \end{aligned}
  \right\}
$,
$
  {
    j = \overline{0, N_0 - 1}
  }
$.
For $k = \overline{1, N - 1}$, construct the optimal solution for $P_k$ from optimal solutions for $P_{k'}$, $k' = \overline{0..k-1}$.
\begin{equation*}
  \left\{
    \begin{aligned}
      & T_{k, j}^{\#}\\
      & T_{k, j}^{\epsilon}
    \end{aligned}
  \right\}
  =
  \min
  _
  {
    \begin{aligned}
      0 \leq k' \wedge k' < k\\
      0 \leq j' \wedge j' < N_{k'}\\
      \functioncheck
      {
        \left(
          \left( k', j' \right),
          \left( k, j \right)
        \right)
      }
    \end{aligned}
  }
  {
    \left(
      \left\{
        \begin{aligned}
          & T_{k', j'}^{\#} + 1\\
          & T_{k', j'}^{\epsilon}
          +
          \epsilon
          _
          {
            \left( k', j' \right),
            \left( k, j \right)
          }
        \end{aligned}
      \right\}
    \right)
  }
  ,
\end{equation*}
where
$
  \epsilon
  _
  {
    \left( k', j' \right),
    \left( k, j \right)
  }
$
is the integral square difference between segment
$
  \left( p_{k', j'}, p_{k, j} \right)
$
and the source polyline from vertex $k'$ till $k$,
$
  \functioncheck
  {
    \left(
      \left( k', j' \right),
      \left( k, j \right)
    \right)
  }
$
is a combination of checks described in the previous sections \ref{sec:TestingSegmentTolerance}, \ref{sec:TestingSegmentEndPoints}, and \ref{sec:TestingZigZag} to check if segment
$
  \left( p_{k', j'}, p_{k, j} \right)
$
can describe the part of the source polyline from vertex $k'$ till $k$.

To reconstruct the optimal solution, it is necessary for
$
  \left\{
    \begin{aligned}
      & T_{k, j}^{\#}\\
      & T_{k, j}^{\epsilon}
    \end{aligned}
  \right\}
$
to store $\left\{ k', j' \right\}$ when the right part is minimal.

The optimal solution is reconstructed from
\begin{equation*}
  \min
  _
  {
    0 \leq j \wedge j < N_{N - 1}
  }
  {
    \left\{
      \begin{aligned}
        & T_{N - 1, j}^{\#}\\
        & T_{N - 1, j}^{\epsilon}
      \end{aligned}
    \right\}
  }
\end{equation*}
by recurrently using stored $\left\{ k', j' \right\}$ values.

\subsection
{
  Optimization
  \label{sec:Optimization}
}

It is possible to significantly reduce the complexity of the algorithm described in the previous section~\ref{sec:CombinatorialApproach} by using the approach described in~\cite{PolylineGeneralizationCombinatorical}.
\begin{equation}
  \begin{aligned}
    &
    \min
    _
    {
      \begin{aligned}
        k_1 \leq k' \wedge k' \leq k_2\\
        0 \leq j' \wedge j' < N_{k'}\\
        \functioncheck
        {
          \left(
            \left( k', j' \right),
            \left( k, j \right)
          \right)
        }
      \end{aligned}
    }
    {
      \left(
        \left\{
          \begin{aligned}
            & T_{k', j'}^{\#} + 1\\
            & T_{k', j'}^{\epsilon}
            +
            \epsilon
            _
            {
              \left( k', j' \right),
              \left( k, j \right)
            }
          \end{aligned}
        \right\}
      \right)
    }
    \gtrapprox
    \\
    \gtrapprox
    &
    \min
    _
    {
      \begin{aligned}
        k_1 \leq k' \wedge k' \leq k_2\\
        0 \leq j' \wedge j' < N_{k'}
      \end{aligned}
    }
    {
      \left(
        \left\{
          \begin{aligned}
            & T_{k', j'}^{\#} + 1\\
            & T_{k', j'}^{\epsilon}
            +
            \epsilon
            _
            {
              \left( k', j' \right),
              \left( k, j \right)
            }
            ^
            {
              \left( k_2 \right)
            }
          \end{aligned}
        \right\}
      \right)
    }
    ,
  \end{aligned}
  \label{eq:OptmizationBaseFormula}
\end{equation}
where
\begin{multline*}
  \epsilon
  _
  {
    \left( k', j' \right),
    \left( k, j \right)
  }
  ^
  {
    \left( k_2 \right)
  }
  =
  \\
  =
  \min
  _
  {
    \begin{aligned}
      0 \leq j_2 \wedge j_2 < N_{k_2}\\
      \functioncheck
      {
        \left(
          \left( k', j' \right),
          \left( k_2, j_2 \right)
        \right)
      }\\
      \functioncheck
      {
        \left(
          \left( k_2, j_2 \right),
          \left( k, j \right)
        \right)
      }
    \end{aligned}
  }
  {
    \left(
      \epsilon
      _
      {
        \left( k', j' \right),
        \left( k_2, j_2 \right)
      }
      +
      \epsilon
      _
      {
        \left( k_2, j_2 \right),
        \left( k, j \right)
      }
    \right)
  }
  .
\end{multline*}

From \eqref{eq:OptmizationBaseFormula}, it follows that
\begin{equation}
  \begin{aligned}
    &
    \min
    _
    {
      \begin{aligned}
        k_1 \leq k' \wedge k' \leq k_2\\
        0 \leq j' \wedge j' < N_{k'}\\
        \functioncheck
        {
          \left(
            \left( k', j' \right),
            \left( k, j \right)
          \right)
        }
      \end{aligned}
    }
    {
      \left(
        \left\{
          \begin{aligned}
            & T_{k', j'}^{\#} + 1\\
            & T_{k', j'}^{\epsilon}
            +
            \epsilon
            _
            {
              \left( k', j' \right),
              \left( k, j \right)
            }
          \end{aligned}
        \right\}
      \right)
    }
    \gtrapprox
    \\
    \gtrapprox
    &
    \min
    _
    {
      \begin{aligned}
        0 \leq j_2 \wedge j_2 < N_{k_2}\\
        \functioncheck
        {
          \left(
            \left( k_2, j_2 \right),
            \left( k, j \right)
          \right)
        }
      \end{aligned}
    }
    {
      \left(
        \left\{
          \begin{aligned}
            & T_{k_2, j_2}^{\#}\\
            & T_{k_2, j_2}^{\epsilon}
            +
            \epsilon
            _
            {
              \left( k_2, j_2 \right),
              \left( k, j \right)
            }
          \end{aligned}
        \right\}
      \right)
    }
  \end{aligned}
  \label{eq:OptimizationFormula1}
\end{equation}
and
\begin{equation}
  \begin{aligned}
    &
    \min
    _
    {
      \begin{aligned}
        k_1 \leq k' \wedge k' \leq k_2\\
        0 \leq j' \wedge j' < N_{k'}\\
        \functioncheck
        {
          \left(
            \left( k', j' \right),
            \left( k, j \right)
          \right)
        }
      \end{aligned}
    }
    {
      \left(
        \left\{
          \begin{aligned}
            & T_{k', j'}^{\#} + 1\\
            & T_{k', j'}^{\epsilon}
            +
            \epsilon
            _
            {
              \left( k', j' \right),
              \left( k, j \right)
            }
          \end{aligned}
        \right\}
      \right)
    }
    \gtrapprox
    \\
    \gtrapprox
    &
    \min
    _
    {
      0 \leq j_1 \wedge j_1 < N_{k_1}
    }
    {
      \left(
        \left\{
          \begin{aligned}
            & T_{k_1, j_1}^{\#}\\
            & T_{k_1, j_1}^{\epsilon}
          \end{aligned}
        \right\}
      \right)
    }
    +
    \\
    &
    +
    \left\{
      \begin{aligned}
        &
        \: \: \: \: \: \: \: \: \: \: \: \: \: \: \: \: \: \: \: \: \: \: \: \: \: \: \: \: \:
        1\\
        &
        \min
        _
        {
          \begin{aligned}
            0 \leq j_2 \wedge j_2 < N_{k_2}\\
            \functioncheck
            {
              \left(
                \left( k_2, j_2 \right),
                \left( k, j \right)
              \right)
            }
          \end{aligned}
        }
        {
          \left(
            \epsilon
            _
            {
              \left( k_2, j_2 \right),
              \left( k, j \right)
            }
          \right)
        }
      \end{aligned}
    \right\}
    .
  \end{aligned}
  \label{eq:OptimizationFormula2}
\end{equation}

The maximum of \eqref{eq:OptimizationFormula1} and \eqref{eq:OptimizationFormula2} can be used to skip checking combinations between vertex $k_1$ and $k_2$.

The inequalities \eqref{eq:OptimizationFormula1} and \eqref{eq:OptimizationFormula2} are approximate due to the use of considered locations. However, this allows finding stricter limitations for the solution inside the interval and simultaneously finding the solution for breaking at vertex $k_2$.

It is possible to construct \eqref{eq:OptimizationFormula1} and \eqref{eq:OptimizationFormula2} with exact inequalities by constructing the optimal solution when the end point is not required to end in the considered location. Similarly, the part from vertex $k_2$ to $\left( k, j \right)$ should not be required to end in considered locations for vertex $k_2$. This is useful when the resultant polyline is required to go through the vertices of the source polyline. However, such an algorithm has a worse compression ratio than the one with the flexibility in joints.

See paper~\cite{PolylineGeneralizationCombinatorical} for further details of this algorithm.

\subsection{Optimal Compression of Closed Polylines}

To find the optimal compression of a closed polyline, it is necessary to know the starting vertex. It is also necessary that the resultant polyline starts and ends in the same vertex. The next algorithm will be used to find the starting vertex and construct a closed resultant polyline.
\begin{enumerate}[label={\arabic*.}, ref={\arabic*}]
  \item \label{enum:ConvexHull} Construct a convex hull for all vertices of the source polyline.
  \item \label{enum:SmallesAngle}  Find the smallest angle of the convex hull polygon.
  \item \label{enum:Reorient}  Take the vertex corresponding to the smallest angle as the starting vertex and reorient the closed polyline to start from that vertex.
  \item Apply the algorithm.
  \item \label{enum:ConstructFirstSolution} From the constructed solution, take one vertex in the middle as the new starting vertex and reorient the closed polyline to start from that vertex.
  \item Apply the algorithm once more, while for the first and the last vertex consider only the location of the previous solution for the middle vertex.
\end{enumerate}

Steps \ref{enum:ConvexHull}, \ref{enum:SmallesAngle}, and \ref{enum:Reorient} are important for a small closed polyline. For the small closed polyline, the resultant polyline is within tolerance of the source polyline, even with suboptimal orientation. As a consequence, without these steps, step \ref{enum:ConstructFirstSolution} may not find the optimal division of the source polyline, leading to a suboptimal solution.

\subsection{Optimal Compression by Straight Segments and Arcs}

This algorithm is extendible to support arcs. The arc passing through considered locations differs from the segment by the necessity to define the radius. Unfortunately, it adds significant complexity to the algorithm. Nevertheless, such an algorithm is possible. There are different ways to fit an arc to a polyline: minimum integral square differences of squares~\cite{ThomasReference2, IchokuReference3}, minimum integral square differences~\cite{RobinsonReference6, Landau4, PaperArcFitting, FittingOfCircularArcsWithO1Complexity, EfficientFittingOfCircularArcs}, minimum deviation, etc. Algorithms with complexity $O{\left( n \right)}$, where $n$ is the number of vertices in the fitted polyline, are not suitable due to the significant increase in complexity. The algorithms with acceptable complexity $O{\left( 1 \right)}$ are~\cite{ThomasReference2, IchokuReference3, FittingOfCircularArcsWithO1Complexity, EfficientFittingOfCircularArcs}; however, algorithms based on integral square differences of squares~\cite{ThomasReference2, IchokuReference3} might break for small arcs and, therefore, are not suitable. Checking that the part of the source polyline is within tolerance, end points, and zigzag will be time-consuming due to complexity $O{\left( n \right)}$.

\section{Analysis of the Algorithm Complexity}

The algorithm contains three steps:
\begin{enumerate}[label={\arabic*.}]
  \item Preprocessing: construction of convex hulls (section~\ref{sec:TestingSegmentTolerance}) and filling arrays for an efficient zigzag test (section~\ref{sec:TestingZigZag}).
  \item Construction of the optimal solution (section~\ref{sec:CombinatorialApproach}).
  \item Reconstruction of the optimal solution (section~\ref{sec:CombinatorialApproach}).
\end{enumerate}

A significant amount of time is spent on constructing an optimal solution. It is difficult to evaluate the complexity described in section~\ref{sec:Optimization}; however, the worst complexity is
\begin{equation}
  O{\left( N^2 \cdot \max_{0 \leq i \wedge i < N}{\left( N_i^2 \right)} \cdot \log{\left( N \right)} \right)}
  .
  \label{eq:WorseComplexity}
\end{equation}

The complexity of the algorithm depends on the type of polyline it processes. It is very difficult to conclude what is the practical complexity of this algorithm. If the optimal polyline does not have segments describing too many vertices of the source polyline, \eqref{eq:WorseComplexity} tends to be
\begin{equation}
  O{\left( N \cdot \max_{0 \leq i \wedge i < N}{\left( N_i^2 \right)} \right)}
  .
  \label{eq:PracticalComplexity}
\end{equation}

\begin{figure} [htb]
  \centering
  \includegraphics[width = \columnwidth, keepaspectratio]{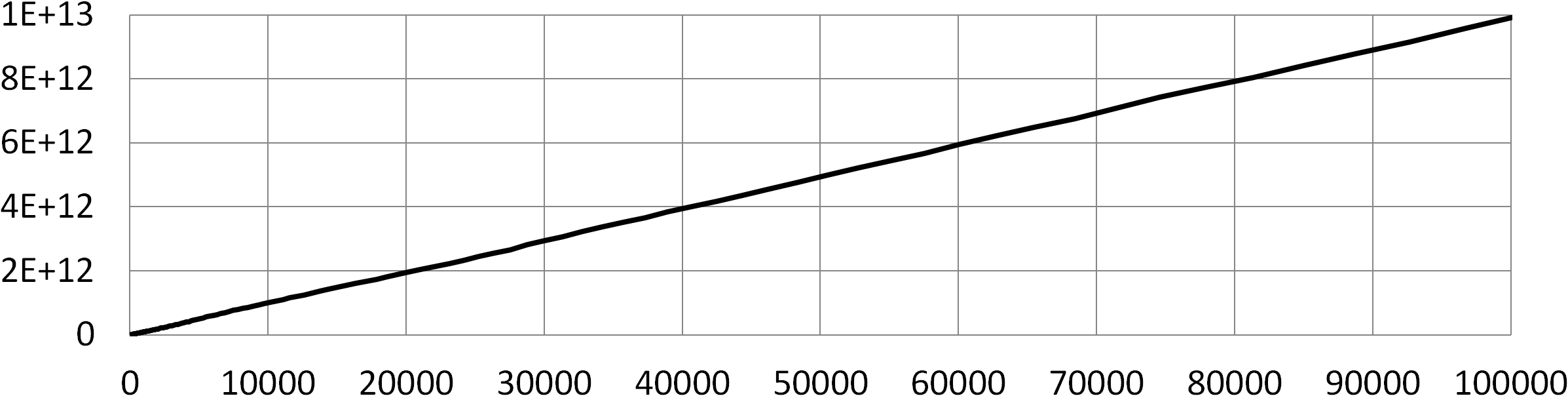}
  \caption
  {
    Time needed to process a polyline versus the number of vertices. The time is measured in CPU ticks on the processor Intel Xeon CPU $\text{E5-2670}$. The polylines are generated by the Brownian motion process. Each next vertex is randomly incremented from the previous vertex by random vector, with components normally distributed with zero mean and $0.25$~standard deviation. The tolerance was set to one. The average reduction in the number of vertices is about $50$~times.
  }
  \label{fig:EstimationOfAlgorithmComplexity}
\end{figure}

Fig.~\ref{fig:EstimationOfAlgorithmComplexity} shows how much time it takes to process a polyline depending on the number of vertices. The dependence is very close to linear, supporting \eqref{eq:PracticalComplexity}.

\section{Examples}

\begin{figure} [htb]
  \centering
  \begin{tabular}{c c}
    \shortstack{(a) \\ \\ \\ \\ (b) \\ \\ \\ \\ (c)} & \includegraphics[width = 7 cm, keepaspectratio]{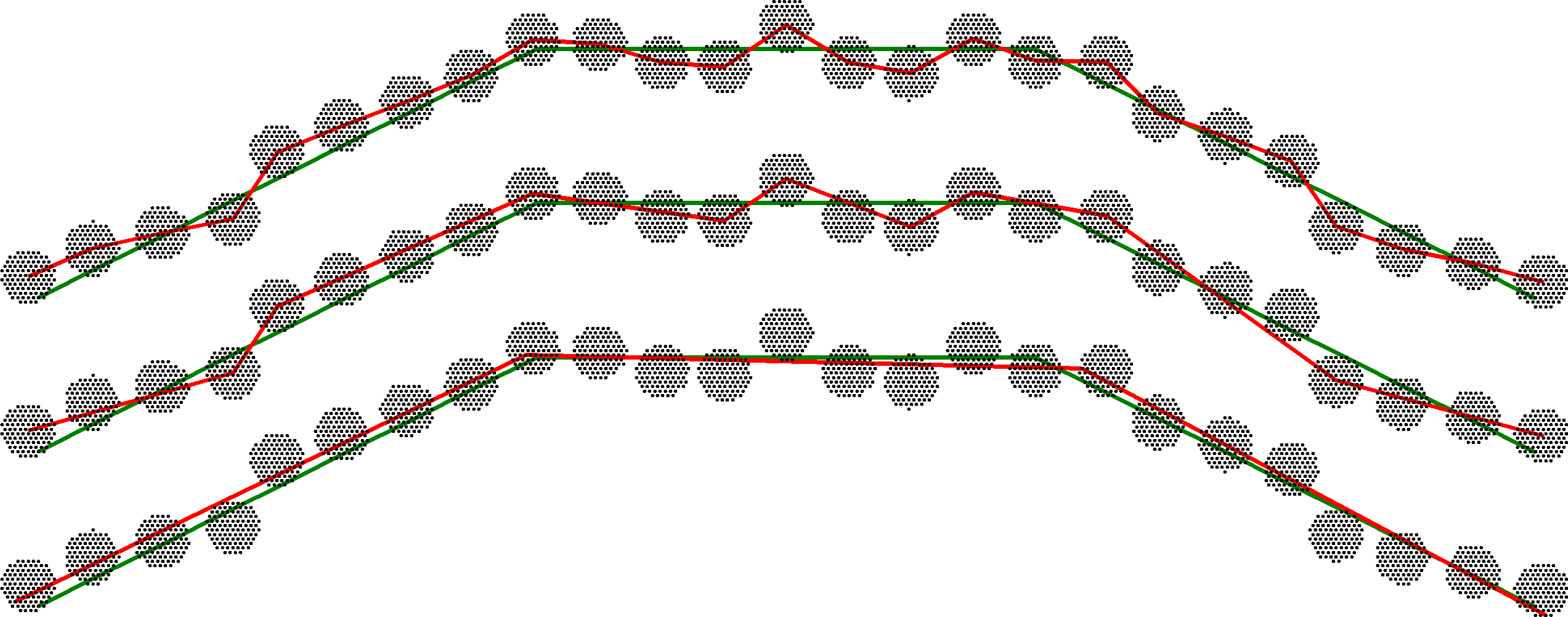}
  \end{tabular}
  \caption
  {
    Comparison of the result of approximation of the Douglas-Peucker algorithm (b) and approximation of optimal polyline compression (c).
    The green polyline is a ground truth. The red polyline is the source polyline (a), the result of the Douglas-Peucker algorithm~\cite{CompressionAlgorithm} (b), and the result of optimal polyline compression (c). The black dots around vertices of the source polyline are considered locations for the vertices of the compressed polyline. The vertices of the source polyline are deviated from the segments of the ground truth polyline by random values uniformly distributed in the interval $\left( -0.1, 0.1 \right)$.
  }
  \label{fig:Comparison}
\end{figure}

Fig.~\ref{fig:Comparison} shows an example of the algorithm described in this paper. If the source polyline is the noisy version of a ground truth polyline, where the noise does not exceed some threshold, and the algorithm is provided with a tolerance slightly greater than the threshold to account for approximations inside the algorithm, then the resultant polyline will never have more vertices than the ground truth polyline.

\begin{figure} [t]
  \centering
    \includegraphics[width = \columnwidth, keepaspectratio]{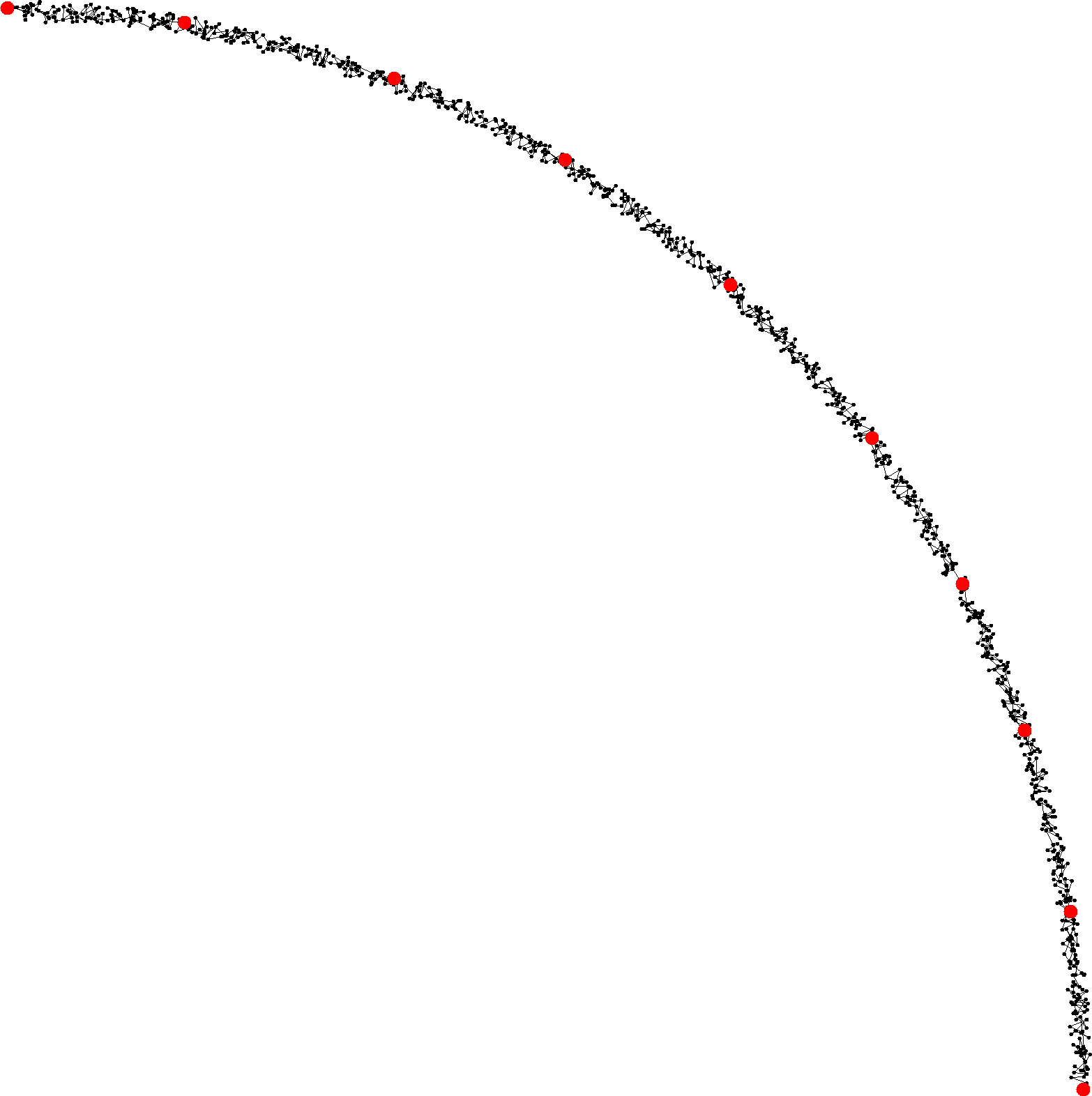}
  \caption
  {
    The black polyline is the source polyline. The red circles are the vertices of the optimal polyline. Ground truth is the arc of $90 \degree$. The noise has uniform distribution in the circle of one percent of the arc radius.
  }
  \label{fig:ExampleArc}
\end{figure}

The effectiveness of the approach is shown in Fig.~\ref{fig:ExampleArc}. Nine segments are sufficient to represent the arc with specified precision. The algorithm not only optimizes the number of segments, it also finds the locations of the segments that minimize integral square differences. Therefore, as shown in Fig.~\ref{fig:ExampleArc}, the algorithm tends to construct segments similar in length.

\begin{figure} [t]
  \centering
    \includegraphics[width = \columnwidth, keepaspectratio]{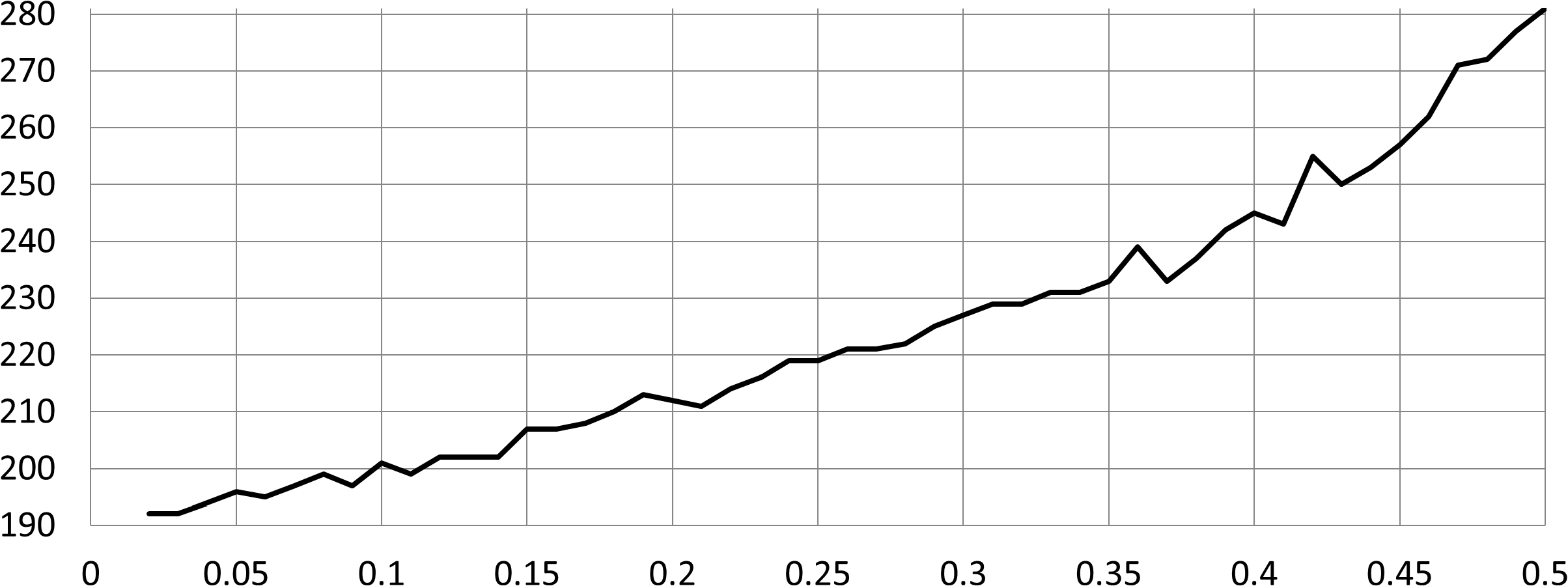}
  \caption
  {
    The number of segments versus discretization error. The polyline was generated by the Brownian motion process in the same way as in Fig.~\ref{fig:EstimationOfAlgorithmComplexity} with $10,000$ vertices.
  }
  \label{fig:GraphCompressionEfficiency}
\end{figure}

Fig.~\ref{fig:GraphCompressionEfficiency} shows the dependence from the error introduced by a discrete set of considered locations (see section~\ref{sec:Descretization}) to the efficiency of the compression. Flexibility in places where neighboring segments connect each other is very important to reach maximum compression, especially for noisy data.

\section{Optimal Compression by Orthogonal Directions}

The triangular grid for considered locations supports directions by $30 \degree$. Reconstruction of orthogonal buildings requires support for $90 \degree$~\cite{ReconstructionOfOrthogonalPolygonalLines} and sometimes $45 \degree$. The square grid for considered locations is more appropriate for this task.

Notice that because only certain directions are allowed, only segments between pairs of considered locations aligned by these directions may be parts of the resultant polyline. Suppose that the resultant segment goes between vertex $i$ and $j$. Because it has to be within tolerance for all vertices between $i$ and $j$, it goes through their considered locations (with the exception of the segment deviating close to the tolerance due to discretization of considered locations).

The optimal solution is found by induction. Define the optimal solution for polyline $P_0$ as
$
  \left\{
    \begin{aligned}
      & T_{0, j, q}^{\#}\\
      & T_{0, j, q}^{\epsilon}
    \end{aligned}
  \right\}
  =
  \left\{
    \begin{aligned}
      0\\
      0
    \end{aligned}
  \right\}
$,
where
$
  j = \overline{0, N_0 - 1}
$,
$
  q = \overline{0, M - 1}
$,
and
$M$ is the number of different directions. For orthogonal case $M = 4$, and for $45 \degree$ case $M = 8$. Take directions as $\alpha_{i} = \dfrac{360 \degree}{M} \cdot i$, $i = \overline{0, M - 1}$.
For $k = \overline{1, N - 1}$, construct the optimal solution for $P_k$ from the optimal solution for $P_{k - 1}$.
\begin{multline*}
    \left\{
      \begin{aligned}
        & T_{k, j, q}^{\#}\\
        & T_{k, j, q}^{\epsilon}
      \end{aligned}
    \right\}
    =
    \\
    \min
    _
    {
      \begin{aligned}
        0 \leq j' \wedge j' < N_{k - 1}\\
        0 \leq q' \wedge q' < M\\
        2 \left| q' - q \right| \neq M\\
        \functionangle{\left( p_{k, j} - p_{k - 1, j'}, \alpha_{q'} \right)}
      \end{aligned}
    }
    {
      \left(
        \left\{
          \begin{aligned}
            & T_{k - 1, j', q'}^{\#} + \delta_{q' \neq q}\\
            & T_{k - 1, j', q'}^{\epsilon}
            +
            \epsilon
            _
            {
              \left( k - 1, j' \right),
              \left( k, j \right)
            }
          \end{aligned}
        \right\}
      \right)
    }
    ,
\end{multline*}
were
$
  \delta_{q' \neq q}
  =
  \left\{
    \begin{aligned}
      1, & \text{ if } q' \neq q,\\
      0, & \text{ otherwise};
    \end{aligned}
  \right.
$
\\
$ \functionangle{ \left( v, \alpha \right)} $ is the check that the vector $v$ has angle $\alpha$ (zero length vectors are allowed).

The condition $2 \left| q' - q \right| \neq M$ corresponds to prohibiting changes in direction by $180 \degree$.

For the $45 \degree$ case, it is possible to restrict the resultant polyline from having sharp angles by not allowing a change of direction by $135 \degree$ ($\left| 4 - \left( \left( q' - q \right) \bmod 8 \right) \right| \neq 1$).

Notice that there are no checks for the tolerance, direction, and end points because they are satisfied during each induction step.

Analyzing the previous solution along $M$ direction will further reduce the amount of calculations. The total complexity of the algorithm is
\begin{equation*}
  O{\left( N \cdot \max_{0 \leq i \wedge i < N}{\left( N_i \right)} \cdot M \right)}
  .
\end{equation*}

For some data, the algorithm may produce an improper result. This happens when the introduction of a zero length segment lowers the penalty.

Because the correct orientation is not known in advance, it is necessary to rotate polylines by different angles and take the solution with the lowest penalty \cite[see section 6]{ReconstructionOfOrthogonalPolygonalLines}.

\begin{figure} [htb]
  \centering
    \includegraphics[width = 6 cm, keepaspectratio]{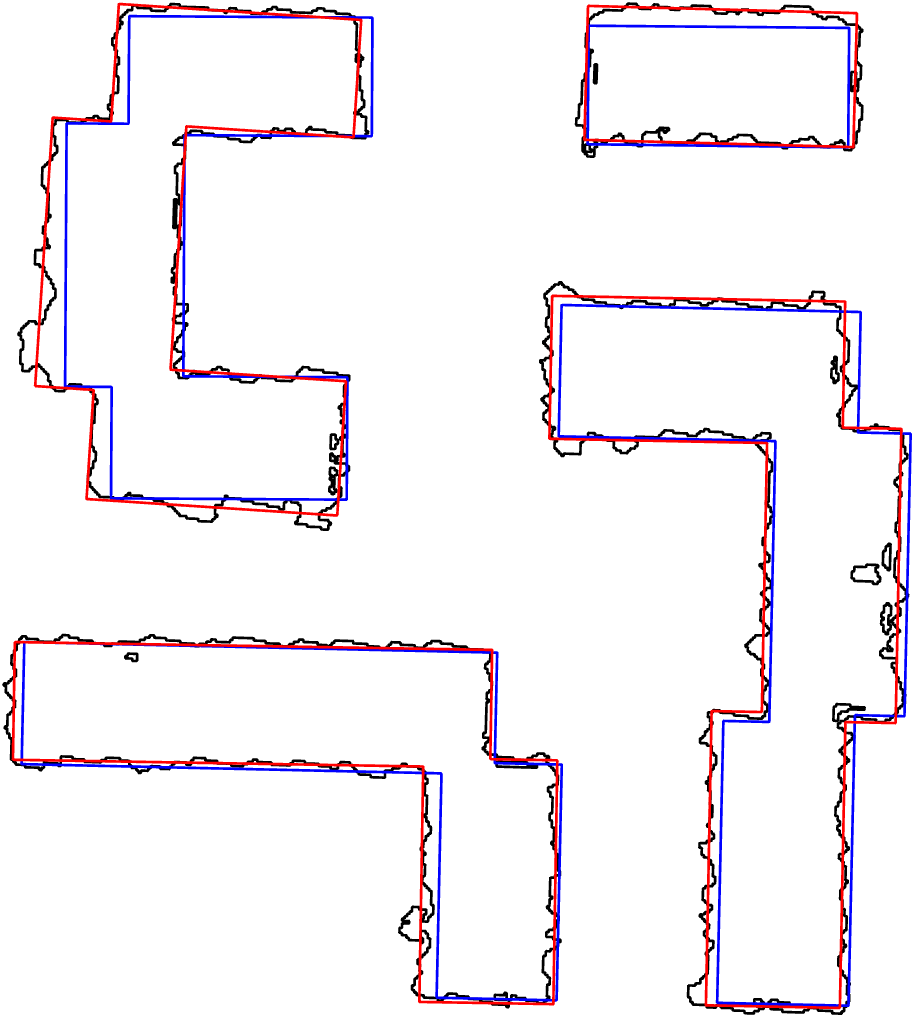}
  \caption
  {
    The black polylines are reconstructed buildings from lidar data \cite{ReferenceLIDARData}. The red polylines are the resultant orthogonal shapes. The blue polylines are the ground truth taken from \cite{ReferenceGroundTruthData}.
  }
  \label{fig:ExampleOrthogonalBuildings}
\end{figure}

Fig.~\ref{fig:ExampleOrthogonalBuildings} shows an example for the reconstruction of orthogonal buildings.

\begin{figure} [htb]
  \centering
    \includegraphics[width = 5 cm, keepaspectratio]{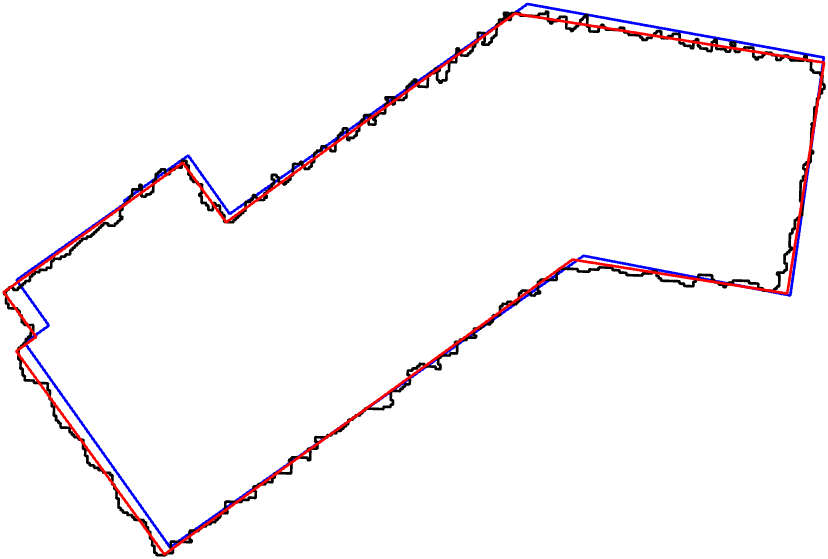}
  \caption
  {
    This differs from Fig.~\ref{fig:ExampleOrthogonalBuildings} by the allowance of $45 \degree$ segments.
  }
  \label{fig:Example45DegreeBuildings}
\end{figure}

The reconstruction of buildings with $45 \degree$ sides are shown in Fig.~\ref{fig:Example45DegreeBuildings}.

The main difference of the algorithm described in this section and \cite{ReconstructionOfOrthogonalPolygonalLines} is in the parameters. The specification of the tolerance is easier than the specification of the penalty $\Delta$ for each additional segment.

\section{Conclusion}

This paper describes an approximation algorithm that finds a polyline with the minimum number of vertices while satisfying tolerance restriction. The solution is optimal with the following limitations:
\begin{itemize}
  \item The vertices of the compressed polyline are limited to considered locations (section~\ref{sec:Descretization}).
  \item The test that the vertex of the compressed polyline is located between some vertices of the source polyline is approximate due to the snapping of the breaking point (section~\ref{sec:Optimization}).
  \item The tests for end points (section~\ref{sec:TestingSegmentEndPoints}) and zigzags are approximate (section~\ref{sec:TestingZigZag}).
\end{itemize}

The performance of the algorithm can be greatly improved if the number of considered locations is decreased without losing quality. This requires further research.

\newcommand{\doi}[1]{\textsc{doi}: \href{http://dx.doi.org/#1}{\nolinkurl{#1}}}


\begingroup
\raggedright
\bibliographystyle{IEEEtran}
\bibliography{CompressedPolyline}
\endgroup


\end{document}